\documentclass[copyright,creativecommons]{eptcs}
\providecommand{\event}{FBTC 2010} 
\providecommand{\volume}{??}
\providecommand{\anno}{20??}
\providecommand{\firstpage}{1}
\providecommand{\eid}{??}
\usepackage{color}
\usepackage{graphicx}
\usepackage{verbatim}
\usepackage{rotating}
\usepackage{subfig}
\usepackage{amsmath}
\usepackage{chemarrow}
\usepackage{comment}
\usepackage{array}

\title{BlenX-based compositional modeling \\ of complex reaction mechanisms}
\author{Judit Z\'{a}mborszky and Corrado Priami
\institute{CoSBi\\ Trento, Italy}
\email{zamborszky@cosbi.eu, priami@cosbi.eu}
}
\def\titlerunning{BlenX-based compositional modeling of complex reaction mechanisms}
\def\authorrunning{J. Z\'{a}mborszky \& C. Priami}

\begin{document}
\maketitle

\begin{abstract}
Molecular interactions are wired in a fascinating way resulting in complex behavior of biological systems. Theoretical modeling provides a useful framework for understanding the dynamics and the function of such networks. The complexity of the biological networks calls for conceptual tools that manage the combinatorial explosion of the set of possible interactions. A suitable conceptual tool to attack complexity is compositionality, already successfully used in the process algebra field to model computer systems. We rely on the BlenX programming language, originated by the beta-binders process calculus, to specify and simulate high-level descriptions of biological circuits. The Gillespie's stochastic framework of BlenX requires the decomposition of phenomenological functions into basic elementary reactions. Systematic unpacking of complex reaction mechanisms into BlenX templates is shown in this study. The estimation/derivation of missing parameters and the challenges emerging from compositional model building in stochastic process algebras are discussed. A biological example on circadian clock is presented as a case study of BlenX compositionality.
\end{abstract}


\section{Introduction}

Computational systems biology is a novel approach to understand how biological systems are orchestrated altogether \cite{Kitano02a,Kitano02b}. Biology, physics, computer science, systems theory and mathematics have joined to propel a multidisciplinary research that provides tools for the analysis of biological studies. Particularly, stochastic approaches are becoming more and more popular as novel experimental techniques - such as quantitative flow cytometry \cite{Darzynkiewicz04} and fluorescence microscopy \cite{DiTalia07} - provide single level measurements of cell physiology. While the average behavior of a cell population has been described by continuous modeling approaches (e.g.\ with Ordinary Differential Equations, ODEs) \cite{Csikasz-Nagy09} from a long time, single cells are analyzed in a stochastic framework as fluctuations may have a significant effect on the physiology of the cell. The influence of noise also in gene expression and signal transduction processes have been shown to be important by both theoretical and experimental approaches \cite{Halling89,McAdams99,Ko92,Levin98,McAdams97,Kierzek01,Ozbudak02,Elowitz02,Rao02}. Computer science through the discipline of Algorithmic Systems Biology \cite{Priami09} enables this research. Here we concentrate on a class of formal languages, namely stochastic process algebras that have been used for interacting and distributed systems and serve as a modeling framework for biological systems as well (for a survey see \cite{Guerriero09}). Starting from stochastic $\pi$-calculus \cite{Priami95,Priami01} and passing through beta-binders \cite{Priami05}, a real programming language has been designed specifically to model and simulate biological systems: BlenX  \cite{Dematte08a}. 

\begin{figure}[!ht]
\centering
\includegraphics[scale=0.2]{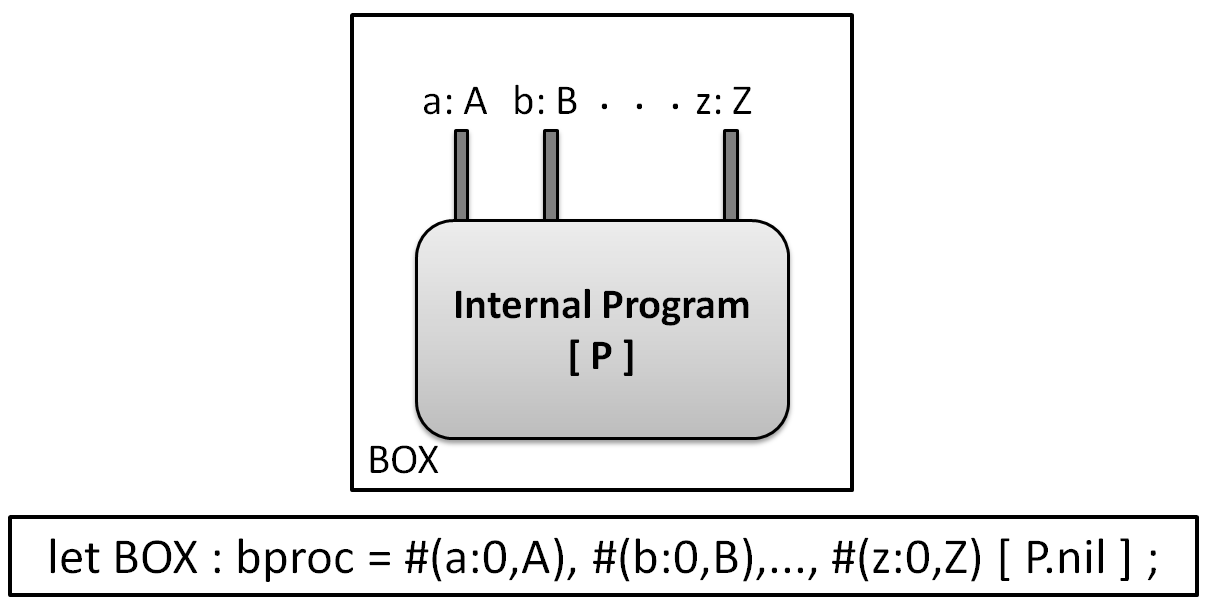}
\caption{Abstraction of biological entities in Blenx.}
\label{box.png}
\label{fig:BOX}
\end{figure}

BlenX allows the modeler to create \textit{boxes} that represent the interacting biological entities (proteins, genes, etc.) (Figure \ref{fig:BOX}). Boxes have an \textit{internal program} (or \textit{internal behavior}) describing their behavior and a set of typed interfaces describing their interaction capabilities. The interaction sites on boxes are called \textit{binders}. The set of binders (\textit{interface}) and the internal program of the box drive its behavior. The behavior of the biological system is given by the ordered sequence of actions and reactions that the program can perform leading to the biochemical interactions between the elements. \textit{Actions} for instance can occur when binders ``sense'' signals (receive an input) and propagate signals (send an output) and the internal structure codifies for the mechanism that transforms an input signal into a change of the box (e.g.\ activation (\textit{unhide} or \textit{expose}), deactivation (\textit{hide}) or changing the type of a binder). Events specify statements to be executed with a rate and/or when some \textit{conditions} are satisfied. Boxes are able to born (when a \textit{new} box is synthesized) and to die (when a box is \textit{deleted}) as biological entities (Table \ref{tab:TABLE1}). Boxes are merged or split upon different conditions (\textit{join} and \textit{split} events). They can also form complexes through their binders and dissociate depending on the state of the overall system. For a more comprehensive description of BlenX we would refer the reader to \cite{Dematte08a}. Similar process calculi initiatives have been also developed for building biological models and performing stochastic simulations (i.e. stochastic $\pi$-calculus \cite{Priami95,Priami01}, BioAmbients \cite{Regev04}, Brane Calculi \cite{Cardelli04}, CCS-R \cite{Danos04a}, $\kappa$-calculus \cite{Danos04b}, Bio-PEPA \cite{Ciocchetta08}.

\begin{table}[!h]
\caption {Basic primitives of BlenX. A declaration of a binder or a box can be nameless or named, e.g.\ it can have an \textit{Id} or \textit{boxId}, respectively. Events are used to express actions that are enabled by global conditions, expressed by \textit{cond}. Actions are that processes can perform.}
\centering
\begin{tabular}{@{}l l l@{}}
\hline\hline
EVENTS: & synthesis of a box & \textbf{when}(\textit{cond or rate}) \textbf{new}(\textit{Decimal}) \\
& degradation of a box	& \textbf{when}(\textit{cond or rate}) \textbf{delete}(\textit{Decimal}) \\
& division of a box	& \textbf{when}(\textit{cond or rate}) \textbf{split}(\textit{boxId,boxId}) \\
& join of a box	& \textbf{when}(\textit{cond or rate}) \textbf{join}(\textit{boxId}) \\
\hline\hline
ACTIONS: & exchange an input/output signal (communication) &	\textbf{a?()}.process  / \textbf{b!()}.process \\
& expose a binder	& \textbf{expose}(\textit{rate, Id:rate, Id}) \\
& hide / show the binder of a box	& \textbf{hide}(\textit{rate,Id}) / \textbf{unhide}(\textit{rate,Id}) \\
& change the type of a binder	& \textbf{ch}(\textit{rate,Id,Id}) \\
\hline
\end{tabular}
\label{tab:TABLE1}
\end{table}

\noindent
Composition is the first challenge that the modelers should deal with. A key innovative aspect of BlenX is the ability to model the reactions between components simply by listing their affinity and without the need of programming all the possible interactions. The BlenX framework allows the user to build systems by fixing each reaction of the network (also called as bottom-up approach) or gives opportunity to handle abstractions as well (such as a top-down approach). After specifying the system, the BlenX program is executed with the Gillespie SSA algorithm \cite{Gillespie77}. The reactions occurring in the system are defined by rate dependent functions that are crucial for the reaction propensities of the stochastic model. Rate functions are associated to actions and events of boxes, and those rates can be determined by the mass action kinetic law. A crucial point of biological models built upon mathematical formalisms is the additional presence of the complex mathematical functions (e.g.\ Michaelis-Menten kinetics \cite{Michaelis13}, Hill function \cite{Hill77}, etc.) that have been empirically developed through several assumptions in order to reduce the size of the system. These abstractions simplify the system leading to a decrease in the required computational power for calculation. Furthermore, modelers often turn to these phenomenological functions to describe the observed behavior of a system without knowing all its details, such as multi-step reactions are often assumed to happen at the same time in cooperative reaction schemes \cite{Weiss97}. Complex rate functions raise several problems in stochastic process algebra approaches.

First, these frameworks are based on Gillespie's stochastic algorithm which considers only elementary reactions, while biological models often deal with nonlinear terms in the deterministic framework. Nonlinearity is known to serve oscillations in several periodic biological systems \cite{Novak08} or multsistability in others \cite{Kholodenko06}, giving an important role for these mathematical formulas in simple models. 

Secondly, when modeling has to deal with several assumptions of phenomenological modules, the freedom of compositionality is reduced. The hidden parts of the modules may contain important linkage between models that are chosen to be merged together. Several biological models are built each day making compositionality to be one of the most important key features of process algebra that offers systematic modeling of complex biological networks. Compositionality has been addressed as an issue of model-construction from elementary reactions with the basic operators \cite{Blossey06,Phillips09}, as the translation of one approach to another \cite{Bortolussi07}, or as the combination of different types of models (ODEs with process algebras, Boolean, hybrid models) \cite{Bortolussi08a}, but the compositionality of complex rate functions has been attracted less attention. 

When non-elementary reactions occur and compound mathematical formulas are used, the direct translation of mathematical terms into the stochastic context is a well-liked approach. Usage of these general functions for calculating the rate of a reaction is also possible in BlenX \cite{Dematte08a,Palmisano08}. However, these implementations have been pointed out by several authors to be incompatible for some cases \cite{Ciocchetta08,Rao03,Mura08,Bundschuh02,Bundschuh03,Paulsson00}, thus modelers have to be careful with them. Stochastic modeling of complex functions is only an approximation and assumptions have to be handled globally. Thus the BlenX framework calls for a semi-automatic method of describing these complex rate functions with intermediate steps (we refer as an ``unpacking'' mechanism) not only owing to ease the compositional programming process, but to provide a correct (and generalized) way of stochastic simulations.

``Unpacked'' version of these modules, representing complex reactions, has to be available as an option for merging them properly when a hidden link is becoming to play a role in the whole system. Usage of decomposed modules containing only elementary steps lengthens simulation time, but provides formal grounds for composition. After merging the modules properly, reduction and abstraction techniques can be applied to the overall system. It is our belief that ``unpacked'' modules could be a natural way of modeling biological processes by connecting small models into large ones. Modern modeling techniques should provide a framework where the model-building process can be carried out correctly (assumptions are taken into account), while it is straightforward and is also suitable for the stochastic simulations. The main goal of this article is to present a possible extension of process algebra tools to carry out compositional modeling in a proper and an easy way.

In this study we concentrate on BlenX. We propose a novel framework for constructing models by previously coded model fragments stored in a template-library. Building large models starting from the basic principles of a process calculus language is time consuming and error-prone, thus a template library should ease the work of programming as it happens in all the fields affected by computer technology. Important motifs with special dynamics have been already proposed in biology \cite{Tyson03,Mangan03} and in computer science \cite{Bortolussi08b}, but they rely on compound phenomenological description and mathematical assumptions. By using BlenX primitives, subsystems (called motifs) are defined, templates are built and decomposition of complex formulas into elementary steps is achieved. Furthermore, compositional modeling is applied to an oscillating biological system (called circadian clock \cite{Roenneberg05}). We show that template-based modeling in BlenX improves compositionality.

The paper is structured as follows. Section 2 and 3 show the stochastic decomposition of Michaelis-Menten kinetics and of Hill functions, respectively. Section 4 uses the stochastic decompositions to compositionally build a model of the circadian clock. We conclude the paper with a discussion of the results and of the future research directions.

\section{Michaelis-Menten module}

Most of the biochemical reactions require catalytic molecules (called enzymes) that increase the rate of the reactions. In enzymatic reactions, the molecules at the beginning of the process are called substrates ($S$), and the enzyme selectively converts them into products ($P$). The kinetic description of such systems was expressed by L. Michaelis and M.L. Menten \cite{Michaelis13}. The derived equation of their results by (referred as Michaelis-Menten kinetics) are widely used in kinetic modeling.
\noindent
The scheme of a one-substrate-one-product reaction (with one active site) is presented in Table \ref{tab:MMreac}. The catalytic step is supposed to be irreversible and the rates of the reactions are given by the law of mass action.

\begin{table}[!h]
\caption{Steps of the enzymatic reaction.}
\centering
\begin{tabular}{l | c l c}
\hline\hline
& (1) & Rate of \textbf{$ES$} complex formation	& $k_1 \cdot E\cdot S$ \\
$E + S \autorightleftharpoons{$k_1$}{$k_2$} ES \xrightarrow{k_3} E + P$ & (2) & Dissociation rate of \textbf{$ES$} complex	& $k_2 \cdot ES$ \\
& (3) & Production rate of \textbf{$P$} &	$k_3 \cdot ES$ \\
\hline
\end{tabular}
\label{tab:MMreac}
\end{table}

\noindent
As enzymes are specific to their substrates and the formation of the enzyme-substrate complex ($ES$) is assumed to be relatively fast, the equilibrium is reached rapidly and the production of $P$ is the limiting step in the overall system. Therefore the $ES$ complex is assumed to be stable, thus the change of its concentration approaches zero (quasi-steady-state assumption (QSSA)). Another important assumption is that the concentration of the substrate highly exceeds the one of the enzyme ($[S] \gg [Etot]$). When a critical substrate concentration is reached, the enzyme is saturated and additional amount of substrate does not influence the velocity of the reaction; it is already maximal ($v_{max}$). 
\noindent
If the last reaction is assumed to be irreversible and all the previously mentioned statements are valid, the rate of the substrate turnover to product is approximated by \\

\begin{tabular}{c c r}
$v=v_{max} \cdot \frac{[S]}{K_m+[S]}$,  &   where $v_{max}=k_3 \cdot [E_{tot}]$, $K_m=(k_3+k_2)/k_1$   and    $[E_{tot}]=E+ES$. 
\end{tabular}

\noindent
This equation provides a complex rate function assuming a single step reaction:    $E + S \xrightarrow{v}E+P$ \\

\noindent
The Michaelis-Menten rate law is often found to be a good approximation to describe enzymatic reactions. The rates of complex formation and its dissociation are rarely available in biological systems, while the key parameters ($v_{max}$ and $K_m$) of a Michaelis-Menten reaction might be easily determined from measured data through linear graphical representations (e.g.\ Lineweaver–Burk plot, Hanes–Woolf plot, Eadie–Hofstee diagram) or by nonlinear regression methods.

In the next subsection, we provide a brief description of how to code enzymatic reations in BlenX with elementary steps and we give a hint how to search for unknown parameters in a Michaelis-Menten module.

\subsection{Decomposition of the module}

The use of Gillespie's stochastic algorithm requires elementary steps instead of complex rate functions in a model. Decomposition of the Michaelis-Menten rate law into elementary reactions may lead to crucial changes in the system's behavior as nonlinearity may disappear if assumptions are inconsistent about enzyme-substrate complexes \cite{Ciliberto07,Sabouri-Ghomi07}. Compositional model building should carefully handle the enzyme molecules hidden in the QSSA.

\begin{figure}[!h]
\centering
\includegraphics[scale=0.3]{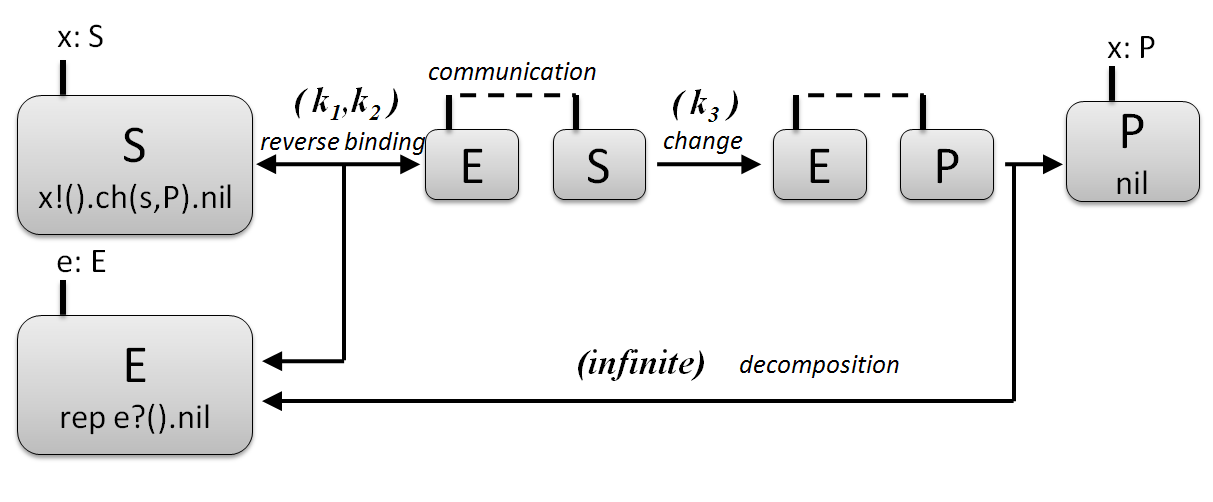}
\caption{BlenX representation of the Michaelis-Menten kinetics.}
\label{mm.png}
\label{fig:MMBlenX}
\end{figure}

\noindent
The Michaelis-Menten module can be implemented easily into BlenX language as the binding of the substrate and the enzyme is described as complex formation through specific binding sites of the boxes representing the proteins. In Figure \ref{fig:MMBlenX}, the types S and E are compatible and equipped with complexation and decomplexation rates. After complex-formation, $S$ and $E$ communicate and the internal behavior of the substrate box is changed into the behavior of the product (the \textit{ch(x,P)} action modifies the type of the binder x into P). The new product has binding affinity no more to the enzyme and an infinite rate decomplexation occurs to release the enzyme $E$.
Decomposition of the nonlinear term into elementary reactions calls for the rate constants of each step. 
\noindent
The rate constant of the catalytic reaction $k_3 = v_{max} /[E_{tot}]$ is easily obtained from the Michaelis-Menten formula. The dissociation constant of the $ES$ complex $k_2 = k_1 \cdot K_m - k_3$ is obtained from the Michealis-Menten constants and it is supposed to be low because the $ES$ complex is assumed to be stable. The rate constants of the reversible complexation ($k_1$ and $k_2$) can be chosen among several combinations by ensuring that the association rate is larger than the dissociation rate of the $ES$ complex. Furthermore, we know that the catalytic step is the rate limiting, thus $k_1$ is chosen to be much larger than $k_3$. Our option determines the time of the simulation, thus values of the rate constants have to be carefully selected. In isolated systems we can scale down the constants easily in order to speed up the simulation. However the rates of the reversible complex formation have to be fast enough and cannot be limiting in a large model.

The proper rate constants describing our compound Michaelis-Menten module have been selected by taking the minimum amount of substrate during the reaction and setting the initial (total) concentration of enzyme to $S_{min} \cdot 0.1$. As a consequence, we get
$
\begin{array}{@{\,\,}l@{\,\,\,}r@{}}
k_1=k_3 \cdot 1000=v_{max}/[E_{tot}] \cdot 1000 & \mbox{and } k_2=K_m \cdot k_1-k_3.
\end{array}
$
\noindent
Note that the selection of feasible parameters must lead to a positive value of $k_2$; and the total concentration of enzyme has to be globally lower than the substrate with a large extent.

\subsection{Simulation results}
\begin{figure}[!h]
\centering
\includegraphics[scale=0.1]{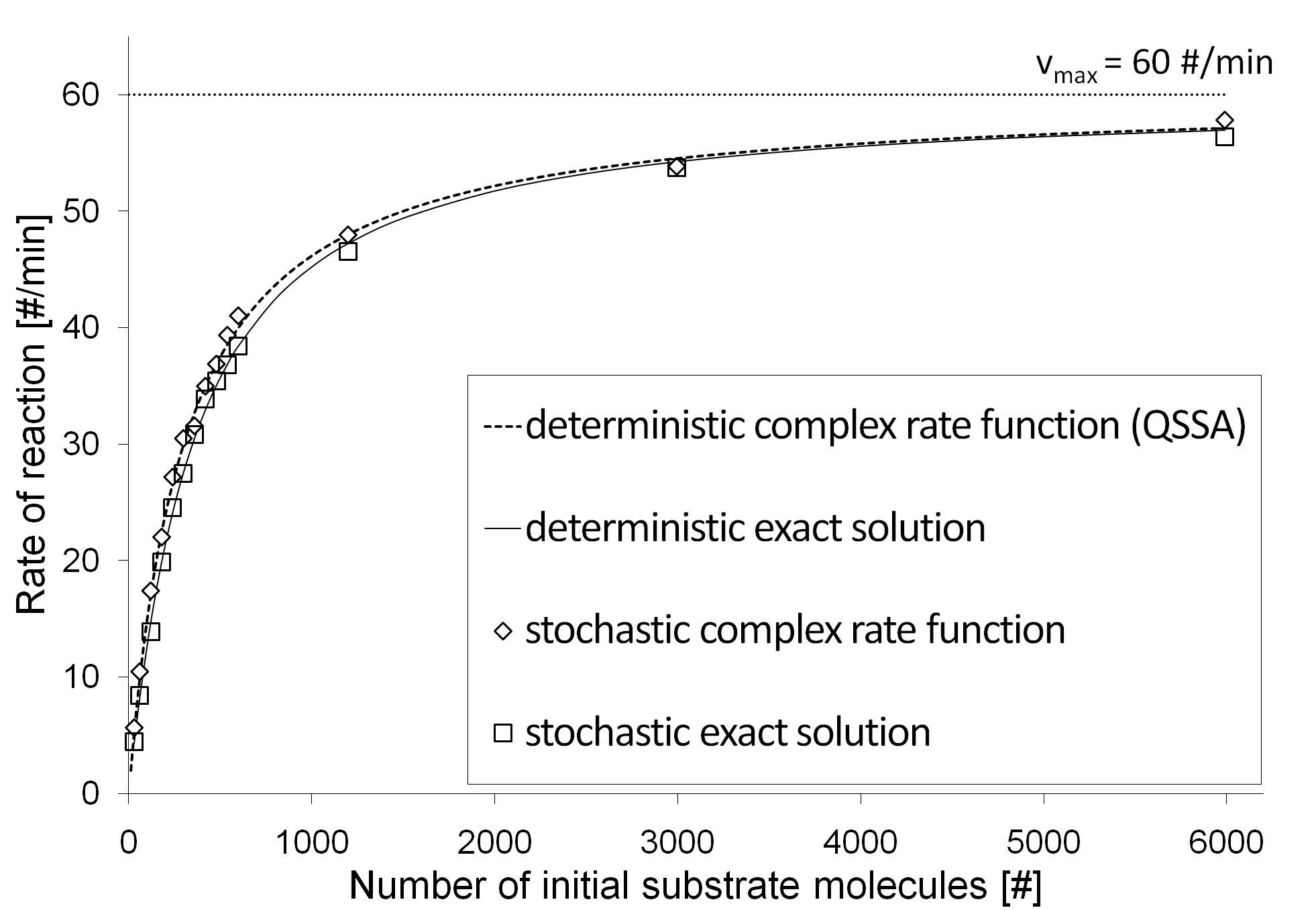}
\caption{Simulation results of the BlenX model fits the deterministic saturation curve well. At low amount of initial substrate molecules the original assumptions of Michaelis and Menten do not match and the exact solution diverges from the complex function.}
\label{fig:200run}
\end{figure}

We analyzed two models with a complex rate function and with an exact solution of the Michaelis-Menten kinetics. First, we converted the concentrations of the deterministic system into molecule numbers through a transformation on the parameters using a scalar constant $\alpha$ defined as $1/(N_A \cdot 10^{-6} \cdot V)$, where $N_A$ is the Avogadro number and $V$ is the volume of the modeled system (see the method of \cite{Palmisano08}). Then we set the models to different initial conditions for the substrate and run $200$ simulations for each initials. The rates of product formation have been derived from the simulation results and these values are plotted over the initial amount of substrate molecules. It provides a saturation curve of the Michaelis-Menten kinetics (Figure \ref{fig:200run}).

\noindent
Calculation of the parameters is based on the assumptions shown previously. $K_m$ is set to $300\#$ ($\#$ refers to number of molecules) and $v_{max}$ is $60\#/min$. The total enzyme amount ($|E_{tot}|$) is set to 60\# molecules and the Michaelis-Menten constants define $k_3$ and the ratio of $k_1$ to $k_2$. The chosen parameters also satisfy the assumption that the value of $k_1$ ($dm^3/(min \cdot mol)$) is much larger than the value of $k_2$ ($1/min$). This condition may be suited by different rates of $k_1$ and $k_2$, although these options only influence the speed of the reaction (and our simulation), but does not change the result (data not shown).

\begin{table}[!h]
\caption{Parameters for the Michaelis-Menten module. $\alpha$ is set to $0.00167$ during the simulations.}
\centering
\begin{tabular}{c | c c}
\hline\hline
\textbf{Parameter names} & \textbf{Parameter values} & \textbf{Parameter units} \\
\hline
$k_1$ & $200 \cdot \alpha$ & $1 / (\textit{min} \cdot \#)$ \\
$k_2$ & $99$ & $1 / \textit{min}$ \\
$k_3$ & $1$ & $1 / \textit{min}$ \\
$K_m$ & $0.5 \cdot \alpha$ & $\#$ \\
$v_{max}$ & $0.1 \cdot \alpha$ & $\# / \textit{min}$ \\
$E$ & $0.1 \cdot \alpha$ & $\#$ \\
\hline
\end{tabular}
\label{tab:MMpar}
\end{table}

\begin{figure}[!h]
\centering
\includegraphics[scale=0.4]{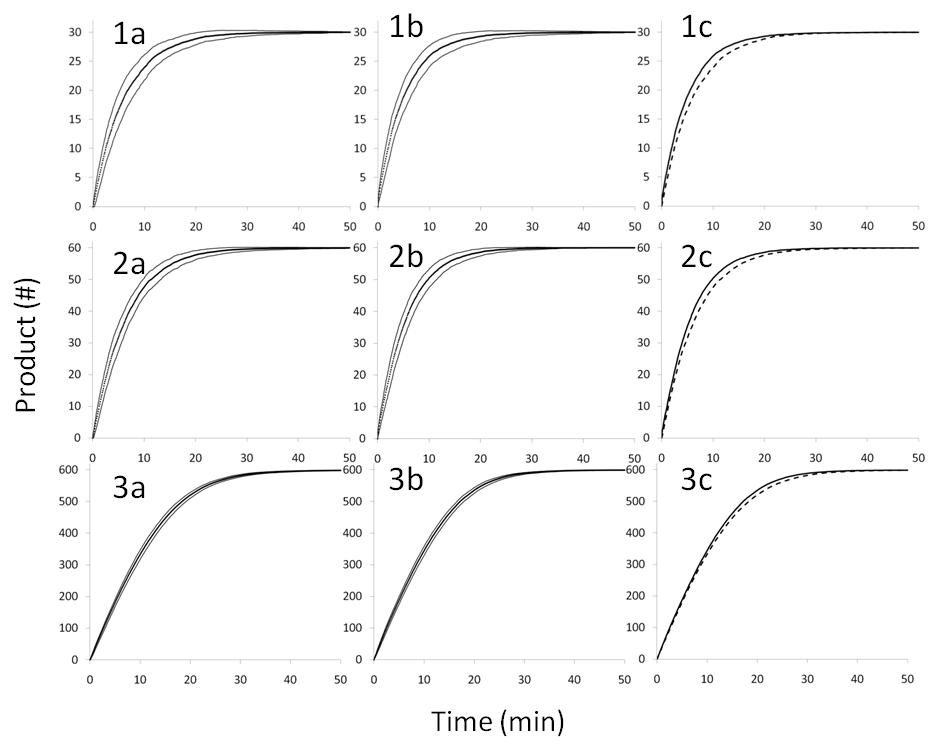}
\caption{Simulation for different number of initial substrates: 1a-c: $S=30\#$; 2a-c: $S=60\#$; 3a-c: $S=599\#$. Average number of product is plotted for each time step. 1-3a: stochastic unpacked versions shown with standard deviation from the mean. 1-3b: complex rate functions with $K_m=300\#$, $v_{max}=60\#/min$ where standard deviation is plotted. 1-3c: comparison of unpacked (dashed) and complex (solid) stochastic simulation results made with BlenX.}
\label{fig:MM}
\end{figure}

\noindent
We compared the deterministic and stochastic simulation results with the ``unpacked'' and complex modules with a parameter set shown in Table \ref{tab:MMpar}. The module built up with complex reaction and the one with elementary reactions shows us a good accordance with each other and also with the deterministic scheme (Figure \ref{fig:200run}). Simulation results of the BlenX model fits the deterministic saturation curve well, although when the original assumptions of Michaelis and Menten do not match, the exact solution diverges from the complex function as it has been shown previously \cite{Rao03}. When the enzyme is in excess of the substrate, the solution of the unpacked model differ greatly from the packed version as the assumption made for the QSSA is not more valid for the system (Figure \ref{fig:200run} and Figure \ref{fig:MM}). This is one limitation of the compound function (also in deterministic simulations).

\noindent
Arkin and Rao assumed \cite{Rao03} that the reactions are isolated and the amount of enzyme is fixed - but in complex network this assumption seems to be ``weak''. Enzyme concentration has to be much less than the substrate concentration and in e.g.\ oscillatory systems the substrate is changed over time. In those cases, the minimum value of the substrate has to provide the base of the calculation (see in Section 4).
We emphasize that decomposition of Michaelis-Menten kinetics is not always necessary, but in a compositional modeling framework it has to be available (as a library). Assumptions have to be checked and the decomposition might be especially useful for further extensions of the model (when hidden details are becoming important). For instance, when an inhibitor of the enzyme is present or two substrates of the same enzyme are introduced, details of the complex reactions have to be elucidated. In Section 2, we provided a brief description of a template for enzyme kinetics in BlenX with a parameter search based on basic mathematical calculus. Implementation of the template-library into the CoSBi Lab platform \cite{CoSBiLab} might automatize the method of parameter estimation as it is a software including inference tools (KInfer \cite{Lecca09}). Research proceed in that direction.

\section{Hill kinetics}
Sigmoidal response curves are usually described by a Hill function. Multiple reactions and several components that form complexes are hidden inside the Hill equation without making the details available. The average of a nonlinear function (e.g.\ Hill function) is generally found to differ from the function of the average \cite{Paulsson00}, thus the proper way of handling these mathematical functions in biological models is crucial. The Hill equation assumes that $n$ molecules of an entity (e.g.\ ligand) bind to a scaffold (e.g.\ receptor) simultaneously \cite{Hill77} and intermediate states do not occur. This is physically possible only if the number of ligands is equal to $1$ ($n=1$), but in most cases it is far from reality.

\begin{figure}[!h]
\centering
\includegraphics[scale=0.2]{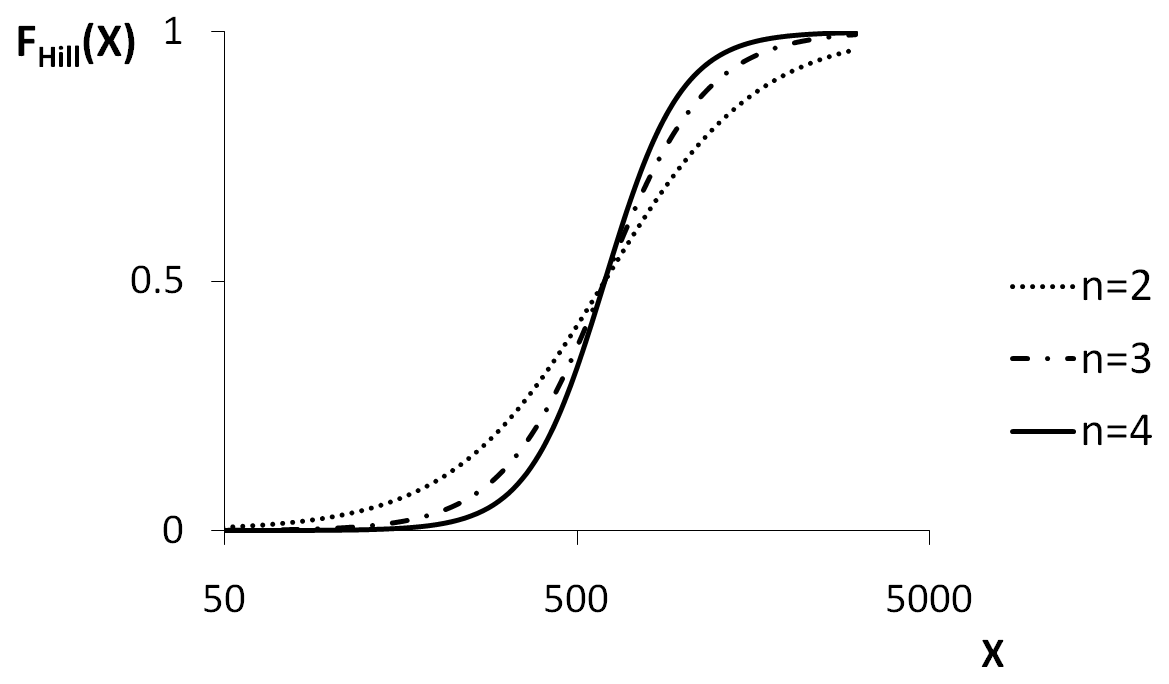}
\caption{Hill function for different Hill coefficients ($n=2,3,4$). When $F_{Hill}(X)$ is $0.5$, $J$ equals to $599\#$.}
\label{FigHill}
\end{figure}

\noindent
Simultaneous binding of $X$ molecules to $Y$ can be described as $Y + nX \autorightleftharpoons{$k_f$}{$k_b$} YX_n$
where $k_f$ is the rate constant of the forward reaction and $k_b$ is of the backward. At equilibrium, the ratio of bound to total receptors is given by the Hill equation
$$F_{Hill} = \frac{Bound}{Total}=\frac{YXn}{Y + YX_n}= \frac{X^n}{X^n + J^n}$$

\noindent
where the dissociation constant is $J = k_b/k_f$ and $n$ provides the number of ligands. The steepness of the transition of the sigmoidal curve depends on the number of ligands ($n$) and $J$ provides the number of ligands at which half of the receptors ($Y$) are bound (Figure \ref{FigHill}). 

\begin{figure}[!h]
\centering
\includegraphics[scale=0.2]{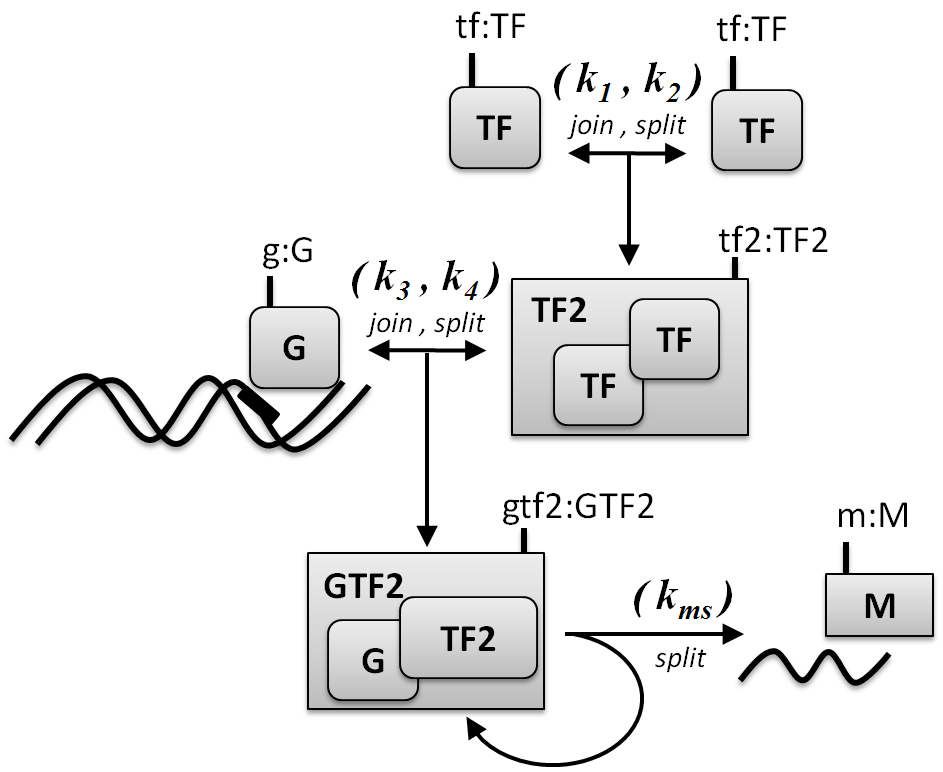}
\caption{Transcriptional regulation: Sequential binding of transcriptional factors ($TF$s) occurs then a homodimer ($TF2$) sits onto the promoter of a gene ($G$) and transcription results in messenger RNAs (mRNAs). Reactions are described through mass action kinetics with rate constants $k_1-k_4$ and $k_{ms}$.}
\label{fig:trans}
\end{figure}

\noindent
Gene expression is known to be a particularly complex and noisy task \cite{McAdams99,McAdams97,Ozbudak02,Elowitz02,Berg78}. Transcriptional regulation is often characterized by a sigmoidal Hill function. Nonlinearity arises from the assumption that the transcription factor forms multimers before binding to DNA (shown in Figure \ref{fig:trans}), creating an abrupt switch between two states.
The detailed mechanism behind the observed behavior is still unclear, but there are several hypothesis and models available \cite{Weiss97,Koshland66,Monod65}. Difficulties of choosing a model has been proposed by \cite{Pedraza08}.
\noindent
In the sequel, we investigate one simplified model of positive cooperativity that captures the requirements of containing only elementary reactions but still maintaining the sigmoidal property of the module. 

\subsection{Decomposition of the module}
Transcriptional factors (abbreviated as $TF$s) often form multimers during transcription \cite{Toledo06} creating sigmoidal response of the system to the change of the transcriptional factors. Cooperativity widely occurs and the Hill equation is a good approximation of the underlying mechanisms, although it assumes simultaneous binding of the $TF$s to the promoter region that is far from the realistic picture. As intermediate states have to occur during the reaction, sequential binding of the transcription factor to the promoter has been considered for this study. The following scheme approximates the Hill function when the intermediate state ($TF2$) does not accumulate and positive cooperativity is present. Several other interactions are plausible \cite{Weiss97}, but are not investigated in this article for the sake of simplicity.

\noindent
BlenX offers a formal and efficient definition of join and split events of boxes as complex formation and decomplexation occur in biological systems. Transcription factors ($TF$s) form multimers ($TF2,3,4…$) through a join event, enhancing the affinity of its binding onto the gene's promoter region ($G$). Positive cooperativity results in a low dissociation constant of the $GTF2$ trimer. The joined complex is able to transcribe mRNAs ($M$s) of the gene through a split event that creates $M$ boxes with the release of the active trimer. Bindings are assumed to be reversible. Elementary reactions of the system are summarized in Table \ref{tab:HillR}.

Sigmoidal curves are often measured in experiments providing specific Hill constants (such as $n$, $J$) to the Hill function. The Hill curve describing e.g.\ a transcriptional regulation scheme is not the proper way to apply stochastic calculations. Elementary steps of the sequential binding scheme contains four rate constants ($k_1$, $k_2$, $k_3$, $k_4$) that have to be determined from the constants $n$ and $J$. Derivation of the missing parameters are calculated from the mass action kinetics describing the system. At equilibrium the intermediate complexes are assumed to be stable, thus

$$\frac{TF2}{TF^2} = \frac{k_1}{k_2}  \mbox{        and         } \frac{GTF2}{G \cdot TF2}=\frac{k_3}{k_4}.$$

\noindent
As the total amount of gene promoters do not change, $G=G_{tot}-GTF2$ leads to the term

$$\frac{GTF2}{G_{tot}} = \frac{TF^2}{\frac{k_2 \cdot k_4}{k_1 \cdot k_3}+TF^2}$$ that is identical to a Hill function 

$$\frac{Bound}{Total}= \frac{X^n}{X^n + J^n}.$$
\noindent
Note that $n=2$ and $J^2=(k_2 \cdot k_4)/(k_1 \cdot k_3)$. The rate of transcription of $M$ is $k_{ms} \cdot TF^2/(J^2+TF^2)=k_{ms}\cdot (GTF2)(G_{tot})$ , where $G_{tot}$ is a constant equals to $1$ in this example.

The Hill coefficient ($n$) and dissociation constant ($J$) determine only the relation of the four rate constants ($J^2=(k_2 \cdot k_4)/(k_1 \cdot k_3)$), but different values may satisfy the reaction scheme. We analyze several choices of $k_1$, $k_2$, $k_3$, $k_4$ and describe them by the response coefficient ($R$) of the sigmoidal curve \cite{Goldbeter81}. The response coefficient allows us to measure the steepness of the transition in this reaction as it has been shown for other cases (such as Goldbeter-Koshland zero-order ultrasensitivity \cite{Goldbeter81}). $R$ coefficient is defined as $S_{0.9}/S_{0.1}$, the ratio of the signal (substrate) amount required to give 90\% saturation relative to the amount required to give 10\% saturation \cite{Koshland66}. The dissociation constant ($J$) is chosen to be equal to $599\#$ and the Hill coefficient ($n$) is $2$. The complex reaction rate is $TF^2/(599^2+TF^2)$. In this case $R = 9$.

\begin{table}[!h]
\caption{Reactions for Hill kinetics for the requirement of at least 2 ligands ($TF$s)}
\centering
\begin{tabular}{l | l c l}
\hline\hline
\textbf{Description} & \textbf{Reactions} & \textbf{Rate constants} & \textbf{Units} \\
\hline
Homodimerization of $TF$ & $2 TF \rightarrow \mbox{TF2}$ & $k_1$ & $1 / (\textit{min} \cdot \#)$ \\
Dissociation of TF2 & $TF2 \rightarrow 2\mbox{TF2}$ & $k_2$ & $1 / \textit{min}$ \\
Formation of an active complex & $TF2 + G \rightarrow \mbox{GTF2}$ & $k_3$ & $1 / (\textit{min} \cdot \#)$ \\
Dissociation of the active complex & $GTF2 \rightarrow \mbox{TF2 + G}$ & $k_4$ & $1 / \textit{min}$ \\
Synthesis of the mRNA & $GTF2 \rightarrow \mbox{M + GTF2}$ & $k_{ms}$ & $1 / \textit{min}$ \\
\hline
\end{tabular}
\label{tab:HillR}
\end{table}

\subsection{Simulation results}
We chose eight different set of parameters (see Table \ref{tab:Hill}) satisfying the following relation: $J^2=(k_2 \cdot k_4)(k_1 \cdot k_3)$. We measured the time average of the bound form ($GTF2$) in case of several levels of initial $TF$ and calculated the response coefficient ($R$) and the actual Hill coefficient ($n'$) for each set of parameters. The derived Hill coefficient equals to $log81/logR$ \cite{Goldbeter81}. The derived dissociation constant ($J'$) is calculated from the points as well as the root mean square error of the fit to the simulation results and the simulation point's error to the theoretical Hill function curve. 

\begin{figure}[!h]
\centering
\includegraphics[scale=0.3]{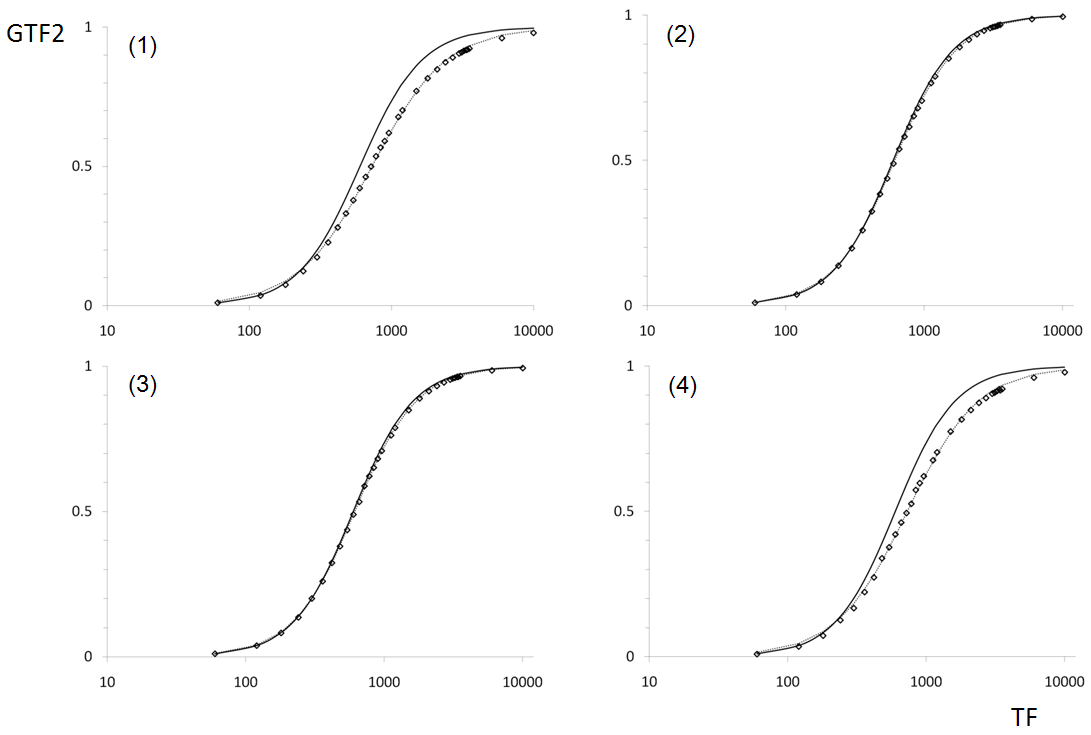}
\caption{The best fits (set 1,2,3,4) satisfy the assumption of positive cooperativity as $K_1 \gg K_2$ and provides a good fit to the theoretical complex function.}
\label{fig:Hillfit}
\end{figure}

If we compare the results of the complex function and the unpacked module, we see that when the assumption of $K_1 \gg K_2$ is valid, the decomposed module gives a good fit to the theoretical Hill curve (Figure \ref{fig:Hillfit}). Our results agree with the observation of \cite{Weiss97} that for simple sequential binding schemes the only condition under which the Hill coefficient does accurately estimate the number of binding sites is when marked positive cooperativity is present. Furthermore, our analysis indicate that the larger the difference between the dissociation constants $K_1$ and $K_2$, the better the fit (e.g.\ compare set $1$ to set $2$). 

\begin{table}[p]
\caption{Multiple simulation results on the module of the Hill kinetics. Set $0$ refers to the deterministic version of the complex Hill function. Set $1$-set $8$ are different sets of parameters for the ``unpacked'' module.}
\centering
\begin{sideways} 
\begin{tabular}{@{}c | c@{\,\,\,\,}c@{\,\,\,\,}c@{\,\,\,\,}c@{\,\,\,\,}c@{\,\,\,\,}c@{\,\,\,\,}c@{\,\,\,\,}c@{\,\,\,\,}c@{\,\,}c@{\,\,\,\,}c@{\,\,\,\,}c@{}}
\hline\hline
 & & & & &$K_1=$ &$K_2=$ &$J=$ & & &Error of the fitting & &Error of the fitting \\
 &$k_1$ &$k_2$ &$k_3$ &$k_4$ &$k_2/k_1$ &$k_4/k_3$ &$\sqrt[n]{K_1 \cdot K_2}$ &$n'$ &$J'$ & of the estimated & $R$ &of the theoretical \\
 & & & & & & & & & &Hill function & &Hill function  \\
 & $1/(\textit{min} \cdot \#)$ & $1 / \textit{min}$ & $1 / (\textit{min} \cdot \#)$ & $1 / \textit{min}$ & $\#$ & $\#$ & $\#$ & - & $\#$ & - & - & - \\
\hline
set$0$ &- &- &- &- &- &- &$1/\alpha$ &$2$ &$599$ &$0$ &$9$ &$0$ \\
set$1$ &$1 \cdot\alpha$ &$10$ &$1000 \cdot\alpha$ &$100$ &$10$ &$0.1$ &$1/\alpha$ &$1.67$ &$729$ &$0.007511$ &$13.89$ &$0.066913$ \\
set$2$ &$1 \cdot\alpha$ &$100$ &$1000 \cdot\alpha$ &$10$ &$100$ &$0.01$ &$1/\alpha$ &$1.95$ &$612$ &$0.001737$ &$9.52$ &$0.008401$ \\
set$3$ &$10 \cdot\alpha$ &$1000$ &$100 \cdot\alpha$ &$1$ &$100$ &$0.01$ &$1/\alpha$ &$1.95$ &$612$ &$0.002859$ &$9.52$ &$0.600704$ \\
set$4$ &$100 \cdot\alpha$ &$1000$ &$10 \cdot\alpha$ &$1$ &$100$ &$0.01$ &$1/\alpha$ &$1.68$ &$731$ &$0.009201$ &$13.68$ &$0.709089$ \\
set$5$ &$10 \cdot\alpha$ &$1$ &$100 \cdot\alpha$ &$1000$ &$0.1$ &$10$ &$1/\alpha$ &$1.05$ &$12418$ &$0.002033$ &$65.71$ &$0.600837$ \\
set$6$ &$1 \cdot\alpha$ &$1$ &$10 \cdot\alpha$ &$1000$ &$0.01$ &$100$ &$1/\alpha$ &$1$ &$124191$ &$0.000234$ &$81.00$ &$0.709229$ \\
set$7$ &$1000 \cdot\alpha$ &$100$ &$1 \cdot\alpha$ &$10$ &$0.1$ &$10$ &$1/\alpha$ &$1.06$ &$12248$ &$0.001434$ &$63.16$ &$0.600704$ \\
set$8$ &$1000 \cdot\alpha$ &$10$ &$1 \cdot\alpha$ &$100$ &$0.01$ &$100$ &$1/\alpha$ &$1.03$ &$110937$ &$0.000255$ &$71.27$ &$0.709089$ \\
\end{tabular}
\end{sideways}
\label{tab:Hill}
\end{table}

We also see that the Hill coefficient is not equal to the number of binding sites on the gene, but provides only a minimum value. 

The simulation results also shows behavior coincident to frequency modulation theory \cite{Pedraza08,Cai08} where we see dense bursts of active transcription factors ($GTF2$) that results burstlike transcription giving similar downstream results that occur in concentration dependent transcription in deterministic models (data not shown).

\section{A biological example}
After laying down the two widely used biological regulatory modules in BlenX, we move further and show how to use them during the composition of a biological case study. We chose circadian clock \cite{Roenneberg05} as an example to present the arising noise of an oscillatory network containing complex rate functions. Circadian clock gives cells a daily rhythm to properly achieve several functions in a 24h periodic manner. This 24h oscillatory system is based upon a negative feedback loop producing a time delayed downregulation of a transcriptional activator. The presented scheme of the clock is adopted from a previous work done by \cite{Zamborszky07}. The model is a simplified model where nonlinearity has an important role in producing robust cycles, thus decomposition of complex rate functions is an interesting task. It is also an oscillating system where the assumptions have to stay valid for the whole model and where the search of parameters is challenging. The proteins present in this system have several roles \cite{Knippschilda05}, thus the possibility of choosing between complex rate functions or elementary reactions containing the required elements (e.g.\ an enzyme) of the reaction explicitly provides a more flexibile use of the language for later extension of the model. Compositionality remains an important and helpful advantage of BlenX in a template based environment. 

\begin{figure}[!h]
\centering
\includegraphics[scale=0.2]{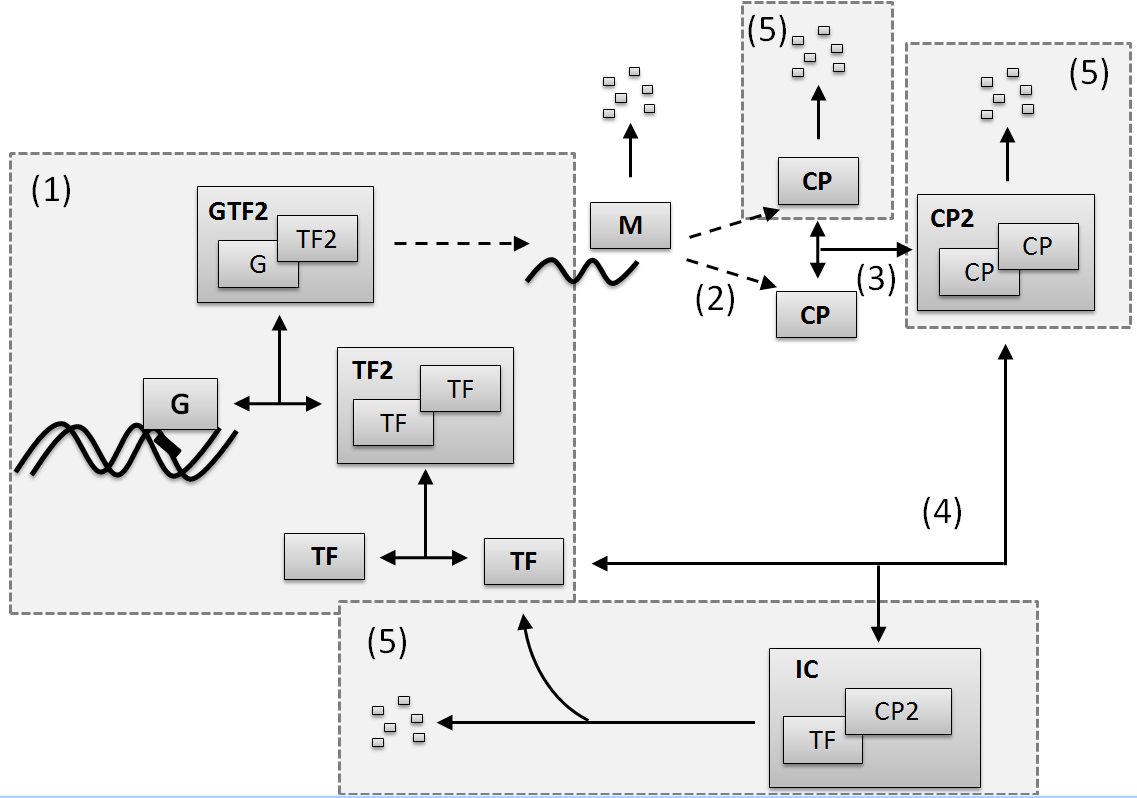}
\caption{Circadian clock model composed of transcriptional and enzyme catalyzed degradation modules.}
\label{fig:circ}
\end{figure}

\begin{figure}[!h]
\centering
\includegraphics[scale=0.3]{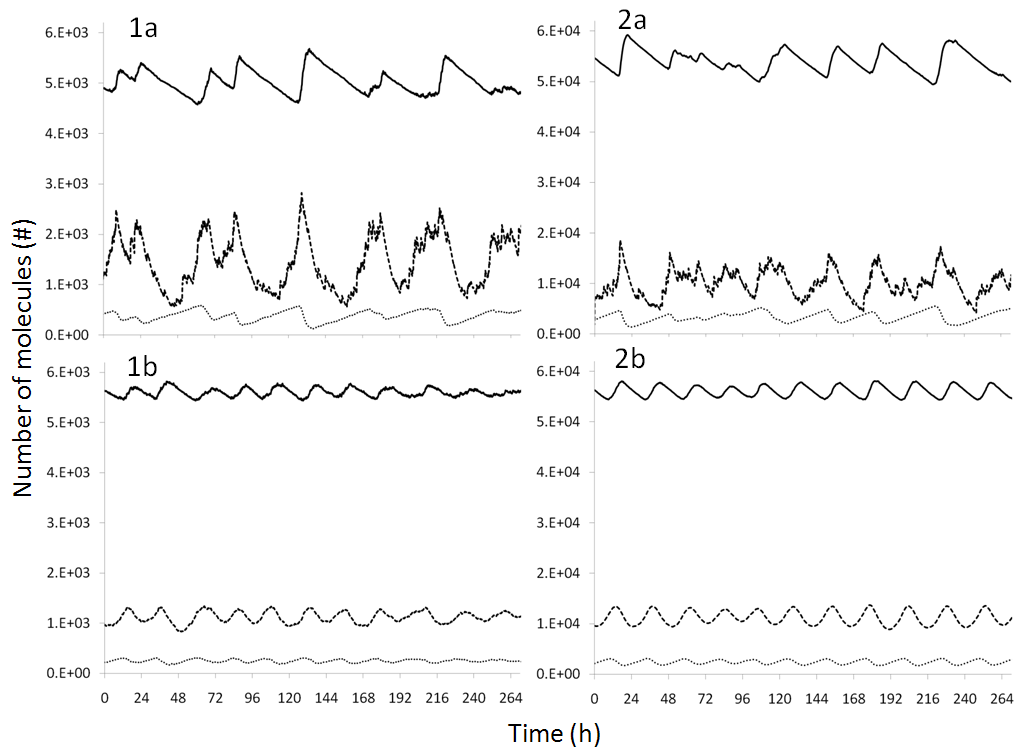}
\caption{Simulation results for the stochastic model containing complex rate functions (1-2b) and for the ``unpacked'' versions (1-2a) in case of a conversion factor $\alpha$=0.000167 (1a-b) and $\alpha$=0.0000167 (2a-b). The total amount of $CP$ in the system is plotted as solid curves; dashed curves represent the messenger ($M$) while dotted points demonstrate the free transcription factors ($TF$) in the model.}
\label{fig:simcirc}
\end{figure}

The system is divided into the following modules: (1) transcriptional regulation following Hill kinetics (2) translation mechanism (assumed to follow mass action kinetics in this study) (3) homodimerization of clock proteins (CP) (4) formation of an inactive complex providing a negative effect inside the loop (5) There are three degradation term catalytically activated by enzymes (following Michaelis-Menten assumption) and the system also contains linear (so called background) degradation of the elements. This network (Figure \ref{fig:circ}) is built with complex rate functions and provides a 24h periodic oscillator (Figure \ref{fig:simcirc} (1-2b)).\vspace{-1em}

We ``unpacked'' the transcriptional step and the three degradation terms following Michaelis-Menten kinetics. We chose the number of enzymes having role in the system to be less than the corresponding substrates, making the assumptions of Michaelis-Menten kinetics valid. The parameters originating from the complex functions are also scaled up to be fast enough. Thus the reactions assumed to be in equilibrium do not limit the system. The products of the enzymatic reactions are degraded immediately (with an infinite rate) after their production in order to serve the catalyzed degradation scheme in the system. 

\begin{figure}[!hp]
\centering
\includegraphics[]{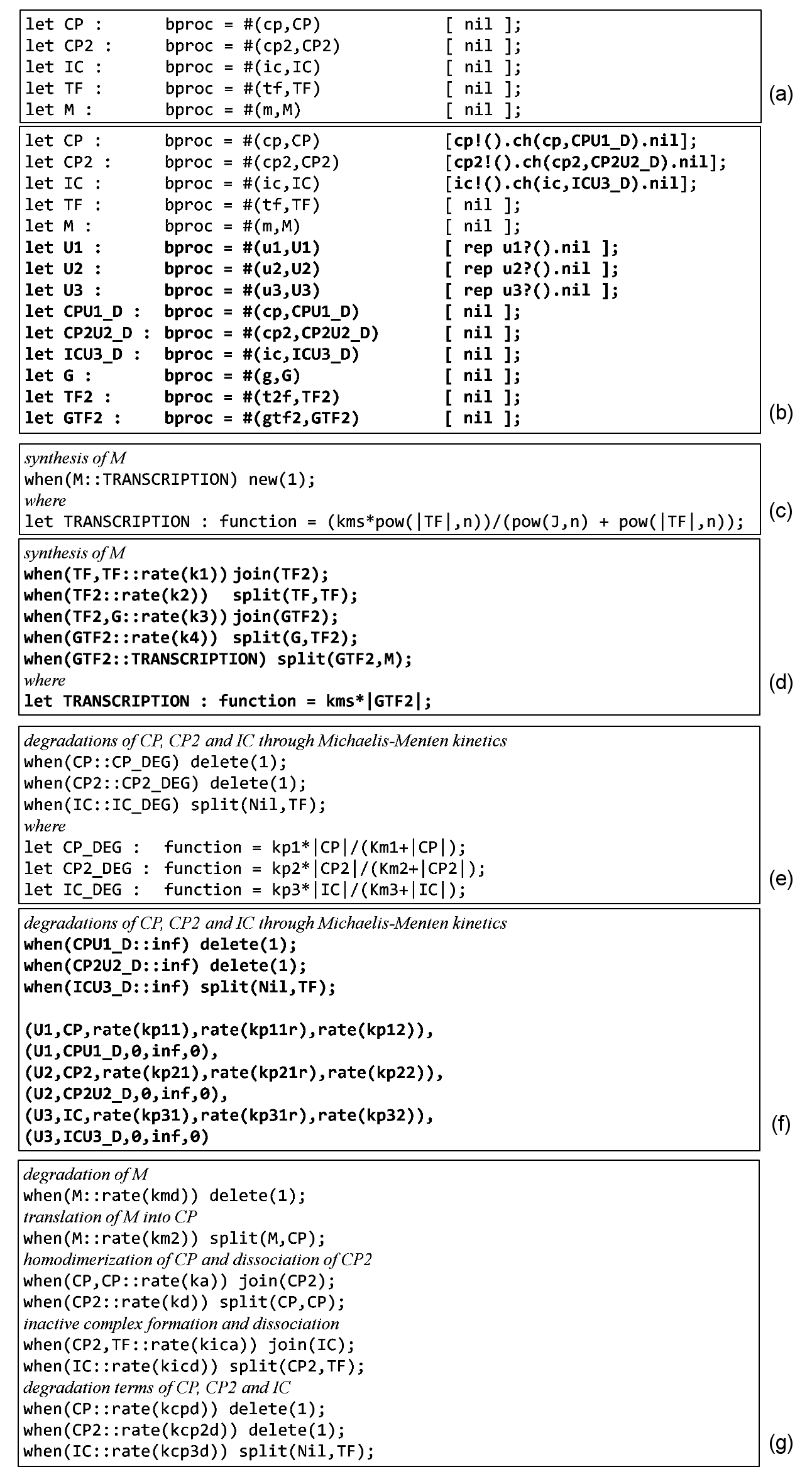}
\caption{BlenX source code. The boxes of the original model are shown in (a) while the ``unpacked'' version is in (b). Composition of (a) and (b) is a straightforward job by parallelization. Events of the original model are in (c), (e), (g), while (d), (f), (g) contains the ``unpacked'' version of the model. Note that there is no change in (g). The substituited modules are highlighted (bold font) in the text. Parameters are provided upon request.}
\label{fig:code}
\end{figure}

We simply merge the modules and insert the boxes (enzymes and intermediates) of the novel entities. We also replace the events corresponding to the complex functions for the ones from the ``unpacked'' modules. This method can be easily automatized as it does not require the modification of the reactions that are independent of the substituted complex functions and the functions calculating the rate of complex reactions does not involve binders. The novel internal behavior of the boxes can be easily parallelized with the original ones. The composition of modules is shown in Figure \ref{fig:code}. Simulation of the ``unpacked'' system (Figure \ref{fig:simcirc} (1-2a)) shows larger noise than the original (Figure \ref{fig:simcirc} (1-2b)), but still produces regular oscillations.
It has been shown in several works \cite{Blossey06,Blossey07} that with process calculus based languages dynamic models can be constructed and existing continuous models can be transferred into the stochastic framework providing additional predictions of the biological system results to the existing models \cite{Mura08}. Herein we have to remind the reader that generally distributed reaction times have been also implemented into the BlenX framework recently \cite{Mura09}. It provides choices of the reaction time distribution for the stochastic simulation algorithm of Gillespie. In this way, abstracted rate laws can be handled stochastically that leads to a better quantitative tool for matching wet-lab experiments and in-silico results without breaking down the complex reactions into elementary steps. The use of this extension fits well the idea of a template based modeling framework as, depending on the question the user asked, biological models might be characterized through complex rate laws and handled by generalized distributions of time; while templates (including only elementary steps) offer a straightforward, flexible and more precise way of compositional modeling in BlenX.

\section{Conclusion}
Two basic complex rate functions that are widely used in biological modeling have been decomposed into elementary reactions in the BlenX framework. The work proposed in this article provides a proper structure for Gillespie's algorithm. Furthermore, the templates offer a straightforward method for compositional modeling. The usage of templates provides a less error prone method in modeling as assumptions are taken into account during the model composition. We exemplified our approach on a circadian clock model. On the top of the first set of predefined templates, a higher level of abstraction and higher level of model composition might be exploited in the future by building larger templates. For instance, a biological switch might be built from two Michaelis-Menten functions, often called Goldbeter-Koshland module \cite{Goldbeter81}. Although the simulation time increases by decomposing complex reactions into elementary ones, we think that the modeling advantage is worth pursuing. Optimizations and simplification can be postponed to the model of the overall system. We will continue investigating the decomposition of complex reactions in order to build a library that will allow drag-and-drop style of modeling complex behavior relying on predefined bricks.
\vspace{2em}

\noindent
\textbf{{\large Acknowledgments}}\\
We thank Attila Csik\'{a}sz-Nagy for the helpful discussions.
\newpage

\section{Bibliography}
\renewcommand*{\refname}{\vspace{-2em}}
\bibliographystyle{eptcs}

\end{document}